\documentclass[preprint,showpacs,preprintnumbers,amsmath,amssymb]{revtex4}
\usepackage{graphicx}

\newcommand{\bg}{\mbox{\boldmath $\gamma$}}
\newcommand{\bt}{\mbox{\boldmath $\theta$}}
\newcommand{\bp}{\mbox{\boldmath $\phi$}}
\begin{document}

\title{Emergent General Relativity from Fisher Information Metric}


\author{Hiroaki Matsueda}
\affiliation{
Sendai National College of Technology, Sendai 989-3128, Japan
}
\date{\today}

\begin{abstract}
We derive the Einstein tensor from the Fisher information metric that is defined by the probability distribution of a statistical mechanical system. We find that the tensor naturally contains essential information of the energy-momentum tensor of a classical scalar field, when the entropy data or the spectrum data of the system are embedded into the classical field as the field strength. Thus, we can regard the Einstein equation as the equation of coarse-grained states for the original microscopic system behind the classical field theory. We make some remarks on quantization of gravity and various quantum-classical correspondences.
\end{abstract}

\maketitle

\section{Introduction}

The power of information-theoretical approaches has been recognized in wide areas of theoretical physics. One of key quantities in quantum field theory is the so-called entanglement entropy. On the other hand, in general relativity and string theory, a long-term topic is to understand the black hole entropy and its microscopic origin, and in recent trends the entropy is almost identified as the entropy of entanglement between the inside and the outside of event horizon~\cite{Solodukhin}. Also in condensed matter physics, it has been attracting considerable attention to examine the scaling properties of the entanglement entropy and its application to the construction of tensor-network wave functions~\cite{Matsueda1}.

Up to now, we have been discussing quantum information itself by using the entanglement entropy. Next, it is natural to extend our analysis to the study of the difference among quantum informations. This is realized by defining the information space or the memory space where the quantum informations are stored by an appropriate format. In this space, an abstract distance should be defined so that the difference between two quantum states is properly measured. A better format for the measure of the difference is to set the coordinate system corresponding to a set of interaction parameters and other key parameters, since they naturally changes the quantum state. Then, the information space seems to have a function similar to the phase diagram. Depending on the essential properties of the states, the information space is a manifold~\cite{Amari}. Then, we focus on geometric properties of the manifold and their close connection to the phase diagram.

In general, the difference among quantum/classical statistical data can be measured by various types of metrices. Well-known examples are Bures metric, relative von-Neumann entropy (Kullback-Leibler divergence), relative Tsallis entropy, and relative R\'{e}nyi entropy. The Bures metric calculates the inner product between two quantum states and subtracts the unity component, while the others measure the difference between two sets of probability distributions (eigenvalue distributions of the density matrix in quantum cases). These abstract distances are globally different, but their local structures are common, except for a constant factor~\cite{Cencov}. The universal metric, or the minimal component, is the so-called Fisher metric. For instance, the Tsallis relative entropy, one of $q$-extention family of the entropy, can be mapped onto the Fisher metric times non-extensive parameter $q$. In a special case, the Fisher metric has the Hessian structure that corresponds to a real version of the K\"{a}hler manifold, and there are beautiful geometric properties~\cite{Shima}. Therefore, if we define a natural connection from the Fisher metric, the emergent classical spacetime spanned by the connection may have some universal physical meaning~\cite{Matsueda2}. The Fisher metric is classical one, but this can also be defined from statistical distribution of quantum systems. Thus, the problem is automatically related to quantum-classical correspondence that has been providing us important questions in many physics fields.

The universality is characterized by the symmetry of the solution of the field equation in a usual classical field theory. In particular, the Einstein equation is the fundamental one in classical curved spacetime. Thus, we would like to examine the nature of the Einstein tensor derived from the Fisher metric. The problems here are whether the Einstein tensor automatically leads to the energy-momentum tensor in a classical field model, and what is the origin of the classical field in the light of the presence of a microscopic model behind this classical theory. We will answer these questions, and propose that the Einstein equation is a kind of the equation of the coarse-grained states in the microscopic model. In general relativity, it is an important problem to know the reason why the Einstein equation has close connection with the first law of thermodynamics~\cite{Padmanabhan,Jacobson,Eling,Verlinde}. The present result sheds new light on this problem.

Before going into detail, I think that it would be helpful for readers to take some comments on the present approach. The readers might be confused for the present scenario that the energy-momentum tensor is 'derived' from more fundamental objects behind the Einstein equation. Usually, the energy-momentum tensor by the real matter field is given in advance, and its spatially non-uniform nature determines our spacetime structure. Then, the metric is only the solution of the Einstein equation, and we may say that the main player is the matter field. In that sense, the present approach is to solve a kind of the inverse problem. Here, we imagine that the quantum informations embedded into the memory space would behave as matter field, since it's possible to take a discription of the data storage that does not follow the linear mesh in general. This abstract or fictitious matter field induces curved space represented by the Fisher metric. If we change our view from physics to information science, the concept of the present approach seems to have indirect connection to that of tangible user interface that aims to remove a fence between information and real materials.

The paper is organized as follows. In Sec. II, we introduce the Fisher metric and examine its general properties. In Sec. III, we derive the Einstein tensor from the Fisher metric. Then, in Sec. IV, we consider some special cases that make the tensor simplified. Then, we find that the entropy data of the original model is stored as a field strength of the classical memory space. In Sec. V, we discuss related topics and summarize our study.

\section{Fisher Metric}

First of all, we develop a general theory for a given family of (classical/quantum) statistical distribution, and after that we will consider the relation between the family and microscopic models. Let us first introduce a continuum model. We represent probability distribution as $p(x;\bt)$ where $x$ denotes the stochastic variable that is supported on a region $X$ and $\bt=\left(\theta^{1},\theta^{2},...\right)$ denotes internal parameters that characterize a microscopic model. For the moment, it is enough to consider that $\bt$ is a set of interaction parameters. The probability distribution obeys the following relation
\begin{eqnarray}
\int_{X}p(x;\bt)dx=1. \label{conservation}
\end{eqnarray}
For later convenience, we abbreviate the expectation value of a distribution function $O(x;\bt)$ by the bracket as
\begin{eqnarray}
\left<\mbox{\boldmath $O$}\right>=\int_{X}p(x;\bt)O(x;\bt)dx, \label{avarage}
\end{eqnarray}
where we omit the index $\bt$ in the bracket and use the bold symbol. In addition, we introduce the spectrum $\gamma(x;\bt)$ as
\begin{eqnarray}
\gamma(x;\bt)=-\ln p(x;\bt). \label{gamma}
\end{eqnarray}
Then, the von Neumann entropy is represented as
\begin{eqnarray}
S(\bt) = -\int_{X}p(x;\bt)\ln p(x;\bt)dx = \int_{X}p(x;\bt)\gamma(x;\bt)dx = \left<\bg\right>. \label{entropy}
\end{eqnarray}
A key quantity in this paper is the so-called Fisher metric $g_{\mu\nu}$ defined by
\begin{eqnarray}
g_{\mu\nu}(\bt)=\int_{X}p(x;\bt)\frac{\partial\gamma(x;\bt)}{\partial\theta^{\mu}}\frac{\partial\gamma(x;\bt)}{\partial\theta^{\nu}}dx=\left<\partial_{\mu}\bg\partial_{\nu}\bg\right>. \label{metric}
\end{eqnarray}
For example, this metric is obtained by calculating the Kullback-Leibler divergence
\begin{eqnarray}
D_{KL} = \int_{X}p(x;\bt)\ln p(x;\bt) dx - \int_{X}p(x;\bt)\ln p(x;\bt+d\bt) dx. \label{KL}
\end{eqnarray}
This is not symmetric for the exchange between $p(x;\bt)$ and $p(x;\bt+d\bt)$, but we find that the second-order expansion has better properties as a measure of the distance. Actually, we can easily derive the following result
\begin{eqnarray}
D_{KL} &\simeq& - \int_{X}p(x;\bt)\left\{ \frac{1}{p(x;\bt)}\frac{\partial p(x;\bt)}{\partial\theta^{\mu}}d\theta^{\mu}-\frac{1}{2}\frac{1}{p^{2}(x;\bt)}\frac{\partial p(x;\bt)}{\partial\theta^{\mu}}\frac{\partial p(x;\bt)}{\partial\theta^{\nu}}d\theta^{\mu}d\theta^{\nu} \right\}dx \nonumber \\
&=& -\frac{\partial}{\partial\theta^{\mu}}\left(\int_{X}p(x;\bt)dx\right)d\theta^{\mu} + \frac{1}{2}g_{\mu\nu}(\bt)d\theta^{\mu}d\theta^{\nu} \nonumber \\
&=& \frac{1}{2}g_{\mu\nu}(\bt)d\theta^{\mu}d\theta^{\nu}.
\end{eqnarray}
In the last equality, we used Eq.~(\ref{conservation}). The physical meaning of the Fisher metric is as follows: When we fix a set of model parameters $\bt$ in a microscopic model, we obtain a particular physical state and its entropy value. Changing $\bt$ infinitesimally, we obtain a bit different physical state and a corresponding entropy value. Collecting all of the entropy data as a function of $\bt$, we can store the data into the classical spacetime with coordinates $\bt$ where the distance is measured by the Fisher metric. The field strength in the classical spacetime is thus the magnitude of the entropy embedded there.

For discrete systems (classical or quantum), we also introduce similar definitions except for some minor corrections. At first, the conservation relation of the probability distribution is given by
\begin{eqnarray}
\sum_{n}p_{n}(\bt)=1. \label{qconservation}
\end{eqnarray}
The expectation value of distribution $O_{n}(\bt)$ is given by
\begin{eqnarray}
\left<\mbox{\boldmath $O$}\right>=\sum_{n}p_{n}(\bt)O_{n}(\bt), \label{qavarage}
\end{eqnarray}
The entroy is given by
\begin{eqnarray}
S(\bt) = -\sum_{n}p_{n}(\bt)\ln p_{n}(\bt) = \sum_{n}p_{n}(\bt)\gamma_{n}(\bt) = \left<\bg\right>, \label{qentropy}
\end{eqnarray}
with the spectrum
\begin{eqnarray}
\gamma_{n}(\bt)=-\ln p_{n}(\bt). \label{qgamma}
\end{eqnarray}
The metric is then represented as
\begin{eqnarray}
g_{\mu\nu}(\bt)=\sum_{n}p_{n}(\bt)\frac{\partial\gamma_{n}(\bt)}{\partial\theta^{\mu}}\frac{\partial\gamma_{n}(\bt)}{\partial\theta^{\nu}}=\left<\partial_{\mu}\bg\partial_{\nu}\bg\right>. \label{qmetric}
\end{eqnarray}
According to Eq.~(\ref{KL}), the Kullback divergence that produces Eq.~(\ref{qmetric}) can be represented by
\begin{eqnarray}
D_{KL} = \sum_{n}\left\{ p_{n}(\bt)\ln p_{n}(\bt) - p_{n}(\bt)\ln p_{n}(\bt+d\bt) \right\}. \label{qKL}
\end{eqnarray}

Here, we need to take care about the presence of Eq.~(\ref{qKL}) in quantum cases, although it is always possible to introduce Eq.~(\ref{qmetric}). The quantum relative entropy, sometimes called Umegaki entropy, is usually defined by the following form
\begin{eqnarray}
D_{U} = {\rm tr}\left(\rho\ln\rho-\rho\ln\sigma\right), \label{Umegaki}
\end{eqnarray}
with use of two density matrices $\rho$ and $\sigma$. In general, their supports $s(\rho)$ and $s(\sigma)$ are different [we assume $s(\rho)\le s(\sigma)$], and we may not find the unitary matrix that simultaneously diagonalizes these matrices. On the other hand, in Eq.~(\ref{qKL}), two probability distributions $p_{n}(\bt)$ and $p_{n}(\bt+d\bt)$ are labeled by the same index $n$, which means the presence of the unitary matrix. Therefore, Eq.~(\ref{qKL}) is a highly restricted relation, if we start from Eq.~(\ref{qKL}). We think that quantum critical systems are safety cases. We can take $\rho$ and $\sigma$ as similar as possible by reducing $d\bt$, while we may not in gapped cases. Thus, the present discussion would be applicable to the critical cases. We will discuss more about this point in the final section. One additional comment is that the parameters $\bt$ are not the bare parameters after the diagonalization. Even the number of the relevant parameters might change by the diagonalization.

For example, let us try to represent geometry of the classical Ising spin model $H=-J\sum_{i,j}\sigma_{i}\sigma_{j}$. Here, $\theta=\beta J$ is the dimensionless internal parameter ($\beta$ is inverse temperature), and we can introduce the Boltzmann distribution
\begin{eqnarray}
p_{n}(\theta)=\frac{1}{Z(\theta)}\exp\left(\theta F_{n}\right), \label{Ising}
\end{eqnarray}
where the index $n$ represents a particular spin configuration $\{\sigma_{1},\sigma_{2},...\}$, $F_{n}=\sum_{i,j}\sigma_{i}\sigma_{j}$, and $Z(\theta)=\sum_{n}e^{\theta F_{n}}$. The Fisher metric is then given by the variance of the quantity $F$
\begin{eqnarray}
g_{\theta\theta}(\theta)=\left<F^{2}\right>-\left<F\right>^{2}=V[F].
\end{eqnarray}
This is energy fluctuation, and thus changes abruptly near the phase transition point $\beta J_{c}$ for spatial dimension larger than one. Therefore, it is natural in this case to imagine that the information space is not simple Euclidean. It is noted that the dimension of the information space is independent of the real spatial dimension of the original Ising model.

The Fisher metric has an another form. Let us start with Eq.~(\ref{conservation}) or Eq.~(\ref{qconservation}), $\left<1\right>=1$. Differentiating both sides of this equation by $\theta^{\nu}$, we obtain
\begin{eqnarray}
\left<\partial_{\nu}\bg\right>=0.
\end{eqnarray}
One more differentiation by $\theta^{\mu}$ leads to
\begin{eqnarray}
\left<\partial_{\mu}\partial_{\nu}\bg\right>=\left<(\partial_{\mu}\bg)(\partial_{\nu}\bg)\right>=g_{\mu\nu}. \label{metric2}
\end{eqnarray}
Thus we have two different representations of the Fisher metric. It should be noted that these two before taking the statistical avarage are not the same
\begin{eqnarray}
\partial_{\mu}\partial_{\nu}\bg\ne(\partial_{\mu}\bg)(\partial_{\nu}\bg).
\end{eqnarray}
Let us further introduce $\mbox{\boldmath $g$}_{\mu\nu}$ as
\begin{eqnarray}
\mbox{\boldmath $g$}_{\mu\nu} = \partial_{\mu}\bg\partial_{\nu}\bg,
\end{eqnarray}
the statistical avarage of $\mbox{\boldmath $g$}_{\mu\nu}$ is equal to the classical one
\begin{eqnarray}
g_{\mu\nu}=\left<\mbox{\boldmath $g$}_{\mu\nu}\right>.
\end{eqnarray}
Thus, $\mbox{\boldmath $g$}_{\mu\nu}$ is a kind of metric, but still contains some fluctuation in the original microscopic model. In the information-geometrical viewpoint, the smooth classical spacetime seems to emerge from this avaraging procedure.

\section{Einstein Tensor of the Fisher Metric}

The purpose of this paper is to calculate the Einstein tensor for the Fisher metric. Thus, let us first calculate the Christoffel symbol defined by
\begin{eqnarray}
\Gamma^{\lambda}_{\;\mu\nu}=\frac{1}{2}g^{\lambda\tau}\left(\partial_{\mu}g_{\nu\tau}+\partial_{\nu}g_{\mu\tau}-\partial_{\tau}g_{\mu\nu}\right), \label{Christoffel}
\end{eqnarray}
where $g^{\lambda\tau}$ is the inverse matrix of the metric. Here, the first derivative of the metric in this symbol is represented by
\begin{eqnarray}
\partial_{\sigma}g_{\mu\nu} = -\left<(\partial_{\sigma}\bg)(\partial_{\mu}\bg)(\partial_{\nu}\bg)\right> + \left<\left(\partial_{\sigma}\partial_{\mu}\bg\right)\partial_{\nu}\bg\right> + \left<\partial_{\mu}\bg\left(\partial_{\sigma}\partial_{\nu}\bg\right)\right>.
\end{eqnarray}
Note that the first term in the right hand side appears as a result of the presence of $p(x;\bt)$ or $p_{n}(\bt)$ in the definition of the bracket. Substituting this into Eq.~(\ref{Christoffel}), most of the terms cancel out, and we obtain
\begin{eqnarray}
\Gamma^{\lambda}_{\;\mu\nu}=g^{\lambda\tau}\left(\left<(\partial_{\mu}\partial_{\nu}\bg)\partial_{\tau}\bg\right>-\frac{1}{2}\left<(\partial_{\mu}\bg)(\partial_{\nu}\bg)(\partial_{\tau}\bg)\right>\right).
\end{eqnarray}
It may be useful for further study to note that the $\alpha$-connection, a more general class of connection, is usually considered in the information geometry. Next, we calculate the Ricci tensor given by
\begin{eqnarray}
R_{\mu\nu} = R^{\sigma}_{\;\mu\sigma\nu} = \partial_{\sigma}\Gamma^{\sigma}_{\;\mu\nu}-\partial_{\nu}\Gamma^{\sigma}_{\;\mu\sigma} + \Gamma^{\sigma}_{\;\rho\sigma}\Gamma^{\rho}_{\;\mu\nu} - \Gamma^{\sigma}_{\;\rho\nu}\Gamma^{\rho}_{\;\mu\sigma}. \label{Riemann}
\end{eqnarray}
Then the Einstein tensor is represented as
\begin{eqnarray}
G_{\mu\nu}=R_{\mu\nu}-\frac{1}{2}g_{\mu\nu}R,
\end{eqnarray}
where the scalar curvature is defined by
\begin{eqnarray}
R=g^{\alpha\beta}R_{\alpha\beta}.
\end{eqnarray}
We devide the right hand side of Eq.~(\ref{Riemann}) into three parts. They are given by
\begin{eqnarray}
R_{\mu\nu} = A_{\mu\nu} + B_{\mu\nu} + C_{\mu\nu}
\end{eqnarray}
where $A_{\mu\nu}$ and $B_{\mu\nu}$ come from the first derivative of the Christoffel symbol, $\partial_{\sigma}\Gamma^{\sigma}_{\;\mu\nu}-\partial_{\nu}\Gamma^{\sigma}_{\;\mu\sigma}$, and $C_{\mu\nu}$ is equal to $\Gamma^{\sigma}_{\;\rho\sigma}\Gamma^{\rho}_{\;\mu\nu} - \Gamma^{\sigma}_{\;\rho\nu}\Gamma^{\rho}_{\;\mu\sigma}$. Their explicit forms are
\begin{eqnarray}
A_{\mu\nu} &=& g^{\sigma\tau}\partial_{\sigma}\left(\left<(\partial_{\mu}\partial_{\nu}\bg)\partial_{\tau}\bg\right>-\frac{1}{2}\left<(\partial_{\mu}\bg)(\partial_{\nu}\bg)(\partial_{\tau}\bg)\right>\right) \nonumber \\
&& - g^{\sigma\tau}\partial_{\nu}\left(\left<(\partial_{\mu}\partial_{\sigma}\bg)\partial_{\tau}\bg\right>-\frac{1}{2}\left<(\partial_{\mu}\bg)(\partial_{\sigma}\bg)(\partial_{\tau}\bg)\right>\right),
\end{eqnarray}
\begin{eqnarray}
B_{\mu\nu} &=& \left(\partial_{\sigma}g^{\sigma\tau}\right)\left(\left<(\partial_{\mu}\partial_{\nu}\bg)\partial_{\tau}\bg\right>-\frac{1}{2}\left<(\partial_{\mu}\bg)(\partial_{\nu}\bg)(\partial_{\tau}\bg)\right>\right) \nonumber \\
&& - \left(\partial_{\nu}g^{\sigma\tau}\right)\left(\left<(\partial_{\mu}\partial_{\sigma}\bg)\partial_{\tau}\bg\right>-\frac{1}{2}\left<(\partial_{\mu}\bg)(\partial_{\sigma}\bg)(\partial_{\tau}\bg)\right>\right),
\end{eqnarray}
and 
\begin{eqnarray}
C_{\mu\nu} &=& g^{\sigma\tau}g^{\rho\zeta}\left(\left<(\partial_{\rho}\partial_{\sigma}\bg)\partial_{\tau}\bg\right>-\frac{1}{2}\left<(\partial_{\rho}\bg)(\partial_{\sigma}\bg)(\partial_{\tau}\bg)\right>\right) \nonumber \\
&& \;\;\;\;\;\;\;\;\cdot\left(\left<(\partial_{\mu}\partial_{\nu}\bg)\partial_{\zeta}\bg\right>-\frac{1}{2}\left<(\partial_{\mu}\bg)(\partial_{\nu}\bg)(\partial_{\zeta}\bg)\right>\right) \nonumber \\
&& - g^{\sigma\tau}g^{\rho\zeta}\left(\left<(\partial_{\rho}\partial_{\nu}\bg)\partial_{\tau}\bg\right>-\frac{1}{2}\left<(\partial_{\rho}\bg)(\partial_{\nu}\bg)(\partial_{\tau}\bg)\right>\right) \nonumber \\
&& \;\;\;\;\;\;\;\;\;\;\cdot\left(\left<(\partial_{\mu}\partial_{\sigma}\bg)\partial_{\zeta}\bg\right>-\frac{1}{2}\left<(\partial_{\mu}\bg)(\partial_{\sigma}\bg)(\partial_{\zeta}\bg)\right>\right).
\end{eqnarray}
In the next step, we will transform the above equations into more compact froms in which their spacetime structures can be seen much better.

First, $A_{\mu\nu}$ is transformed into
\begin{eqnarray}
A_{\mu\nu} &=& g^{\sigma\tau}\left[-\frac{1}{2}\left<(\partial_{\sigma}\bg)(\partial_{\mu}\partial_{\nu}\bg)\partial_{\tau}\bg\right> + \left<(\partial_{\mu}\partial_{\nu}\bg)(\partial_{\sigma}\partial_{\tau}\bg)\right> \right.\nonumber \\
&& \;\;\;\;\;\;\;\; + \frac{1}{2}\left<(\partial_{\nu}\bg)(\partial_{\mu}\partial_{\sigma}\bg)\partial_{\tau}\bg\right> - \left<(\partial_{\mu}\partial_{\sigma}\bg)(\partial_{\nu}\partial_{\tau}\bg)\right> \nonumber \\
&& \;\;\;\;\;\;\;\; \left. - \frac{1}{2}\left<(\partial_{\mu}\bg)(\partial_{\nu}\bg)\partial_{\sigma}\partial_{\tau}\bg\right> + \frac{1}{2}\left<(\partial_{\mu}\bg)(\partial_{\sigma}\bg)\partial_{\nu}\partial_{\tau}\bg\right>\right].
\end{eqnarray}
Then we introduce a part of the Einstein tensor coming from $A_{\mu\nu}$ ($A=g^{\mu\nu}A_{\mu\nu}$) as
\begin{eqnarray}
G_{\mu\nu}^{A} &=& A_{\mu\nu}-\frac{1}{2}g_{\mu\nu}A \nonumber \\
&=& \left<\mbox{\boldmath $\omega$}\left(\partial_{\mu}\partial_{\nu}\bg-\frac{1}{2}g_{\mu\nu}g^{\alpha\beta}\partial_{\alpha}\partial_{\beta}\bg\right)\right> \nonumber \\
&& - \frac{1}{2}g^{\sigma\tau}\left<\partial_{\sigma}\partial_{\tau}\bg\left(\mbox{\boldmath $g$}_{\mu\nu}-\frac{1}{2}g_{\mu\nu}g^{\alpha\beta}\mbox{\boldmath $g$}_{\alpha\beta}\right)\right> \nonumber \\
&& - g^{\sigma\tau}\left<(\partial_{\mu}\partial_{\sigma}\bg)(\partial_{\nu}\partial_{\tau}\bg)-\frac{1}{2}g_{\mu\nu}g^{\alpha\beta}(\partial_{\alpha}\partial_{\sigma}\bg)(\partial_{\beta}\partial_{\tau}\bg)\right> \nonumber \\
&& + \frac{1}{2}g^{\sigma\tau}\left<\mbox{\boldmath $g$}_{\sigma\mu}(\partial_{\nu}\partial_{\tau}\bg)+\mbox{\boldmath $g$}_{\sigma\nu}(\partial_{\mu}\partial_{\tau}\bg)-g_{\mu\nu}g^{\alpha\beta}\mbox{\boldmath $g$}_{\sigma\alpha}(\partial_{\beta}\partial_{\tau}\bg)\right>, \label{GA}
\end{eqnarray}
where $\mbox{\boldmath $\omega$}$ is defined by
\begin{eqnarray}
\mbox{\boldmath $\omega$} = g^{\sigma\tau}\left(\partial_{\sigma}\partial_{\tau}\bg-\frac{1}{2}\mbox{\boldmath $g$}_{\sigma\tau}\right).
\end{eqnarray}
The avarage of $\mbox{\boldmath $\omega$}$ is evaluated as
\begin{eqnarray}
\left<\mbox{\boldmath $\omega$}\right> = g^{\sigma\tau}\left<\partial_{\sigma}\partial_{\tau}\bg\right>-\frac{1}{2}g^{\sigma\tau}g_{\sigma\tau} = \frac{1}{2}D,
\end{eqnarray}
where $D$ is the dimension of our classical space. Next, $B_{\mu\nu}$ is evaluated as
\begin{eqnarray}
B_{\mu\nu} &=& - g^{\sigma\tau}g^{\rho\zeta}\left(-\left<(\partial_{\sigma}\bg)(\partial_{\tau}\bg)\partial_{\rho}\bg\right>+\left<(\partial_{\sigma}\partial_{\tau}\bg)\partial_{\rho}\bg\right>+\left<\partial_{\tau}\bg(\partial_{\sigma}\partial_{\rho}\bg)\right>\right) \nonumber \\
&& \;\;\;\;\;\;\;\;\;\;\cdot\left(\left<(\partial_{\mu}\partial_{\nu}\bg)\partial_{\zeta}\bg\right>-\frac{1}{2}\left<(\partial_{\mu}\bg)(\partial_{\nu}\bg)(\partial_{\zeta}\bg)\right>\right) \nonumber \\
&& + g^{\sigma\tau}g^{\rho\zeta}\left(-\left<(\partial_{\nu}\bg)(\partial_{\tau}\bg)(\partial_{\rho}\bg)\right>+\left<(\partial_{\nu}\partial_{\tau}\bg)\partial_{\rho}\bg\right>+\left<\partial_{\tau}\bg(\partial_{\nu}\partial_{\rho}\bg)\right>\right) \nonumber \\
&& \;\;\;\;\;\;\;\;\;\;\cdot\left(\left<(\partial_{\mu}\partial_{\sigma}\bg)\partial_{\zeta}\bg\right>-\frac{1}{2}\left<(\partial_{\mu}\bg)(\partial_{\sigma}\bg)(\partial_{\zeta}\bg)\right>\right),
\end{eqnarray}
where we have used the following relation
\begin{eqnarray}
\partial_{\sigma}g^{\mu\nu}=-g^{\mu\alpha}g^{\nu\beta}\partial_{\sigma}g_{\alpha\beta}.
\end{eqnarray}
Adding $B_{\mu\nu}$ to $C_{\mu\nu}$, some terms cancel out, and then we obtain
\begin{eqnarray}
B_{\mu\nu}+C_{\mu\nu} &=& g^{\sigma\tau}g^{\rho\zeta}\left(\left<(\partial_{\mu}\partial_{\sigma}\bg)\partial_{\zeta}\bg\right>-\frac{1}{2}\left<(\partial_{\mu}\bg)(\partial_{\sigma}\bg)\partial_{\zeta}\bg\right>\right) \nonumber \\
&& \;\;\;\;\;\;\;\;\cdot\left(\left<(\partial_{\nu}\partial_{\tau}\bg)\partial_{\rho}\bg\right>-\frac{1}{2}\left<(\partial_{\nu}\bg)(\partial_{\tau}\bg)(\partial_{\rho}\bg)\right>\right) \nonumber \\
&& - g^{\sigma\tau}g^{\rho\zeta}\left(\left<(\partial_{\sigma}\partial_{\tau}\bg)\partial_{\rho}\bg\right>-\frac{1}{2}\left<(\partial_{\sigma}\bg)(\partial_{\tau}\bg)(\partial_{\rho}\bg)\right>\right) \nonumber \\
&& \;\;\;\;\;\;\;\;\;\cdot\left(\left<(\partial_{\mu}\partial_{\nu}\bg)\partial_{\zeta}\bg\right>-\frac{1}{2}\left<(\partial_{\mu}\bg)(\partial_{\nu}\bg)(\partial_{\zeta}\bg)\right>\right).
\end{eqnarray}
The above equation can be represented by the following compact form
\begin{eqnarray}
B_{\mu\nu}+C_{\mu\nu} &=& \left<\bp\left(\partial_{\mu}\partial_{\nu}\bg-\frac{1}{2}\mbox{\boldmath $g$}_{\mu\nu}\right)\right> \nonumber \\
&& + g^{\sigma\tau}g^{\rho\zeta}\left<\partial_{\zeta}\bg\left(\partial_{\mu}\partial_{\sigma}\bg-\frac{1}{2}\mbox{\boldmath $g$}_{\mu\sigma}\right)\right>\left<\partial_{\rho}\bg\left(\partial_{\nu}\partial_{\tau}\bg-\frac{1}{2}\mbox{\boldmath $g$}_{\nu\tau}\right)\right>,
\end{eqnarray}
where $\bp$ is defined by
\begin{eqnarray}
\bp = -g^{\sigma\tau}g^{\rho\zeta}\partial_{\zeta}\bg\left<\partial_{\rho}\bg\left(\partial_{\sigma}\partial_{\tau}\bg-\frac{1}{2}\mbox{\boldmath $g$}_{\sigma\tau}\right)\right>.
\end{eqnarray}
The avarage of $\bp$ vanishes, since we have
\begin{eqnarray}
\left<\bp\right> = -g^{\sigma\tau}g^{\rho\zeta}\left<\partial_{\zeta}\bg\right>\left<\partial_{\rho}\bg\left(\partial_{\sigma}\partial_{\tau}\bg-\frac{1}{2}\mbox{\boldmath $g$}_{\sigma\tau}\right)\right> = 0.
\end{eqnarray}
The part of the Einstein tensor coming from $B_{\mu\nu}+C_{\mu\nu}$ is given by
\begin{eqnarray}
G_{\mu\nu}^{B}+G_{\mu\nu}^{C} &=& (B_{\mu\nu}+C_{\mu\nu})-\frac{1}{2}g_{\mu\nu}(B+C) \nonumber \\
&=& \left<\bp\left(\partial_{\mu}\partial_{\nu}\bg-\frac{1}{2}g_{\mu\nu}g^{\alpha\beta}\partial_{\alpha}\partial_{\beta}\bg\right)\right> - \frac{1}{2}\left<\bp\left(\mbox{\boldmath $g$}_{\mu\nu}-\frac{1}{2}g_{\mu\nu}g^{\alpha\beta}\mbox{\boldmath $g$}_{\alpha\beta}\right)\right> \nonumber \\
&& + g^{\sigma\tau}g^{\rho\zeta}\left<\partial_{\zeta}\bg\left(\partial_{\mu}\partial_{\sigma}\bg-\frac{1}{2}\mbox{\boldmath $g$}_{\mu\sigma}\right)\right>\left<\partial_{\rho}\bg\left(\partial_{\nu}\partial_{\tau}\bg-\frac{1}{2}\mbox{\boldmath $g$}_{\nu\tau}\right)\right> \nonumber \\
&& - \frac{1}{2}g_{\mu\nu}g^{\alpha\beta}g^{\sigma\tau}g^{\rho\zeta}\left<\partial_{\zeta}\bg\left(\partial_{\alpha}\partial_{\sigma}\bg-\frac{1}{2}\mbox{\boldmath $g$}_{\alpha\sigma}\right)\right>\left<\partial_{\rho}\bg\left(\partial_{\beta}\partial_{\tau}\bg-\frac{1}{2}\mbox{\boldmath $g$}_{\beta\tau}\right)\right>.
\end{eqnarray}
where $B=g^{\alpha\beta}B_{\alpha\beta}$ and $C=g^{\alpha\beta}C_{\alpha\beta}$.

Finally, the Einstein tensor for the Fisher metric is exactly given by
\begin{eqnarray}
G_{\mu\nu} &=& G_{\mu\nu}^{A}+G_{\mu\nu}^{B}+G_{\mu\nu}^{C} \nonumber \\
&=& \left<\left(\mbox{\boldmath $\omega$}+\bp\right)\left(\partial_{\mu}\partial_{\nu}\bg-\frac{1}{2}g_{\mu\nu}g^{\alpha\beta}\partial_{\alpha}\partial_{\beta}\bg\right)\right> \nonumber \\
&& -\frac{1}{2}\left<\left(\bp+g^{\sigma\tau}\partial_{\sigma}\partial_{\tau}\bg\right)\left(\mbox{\boldmath $g$}_{\mu\nu}-\frac{1}{2}g_{\mu\nu}g^{\alpha\beta}\mbox{\boldmath $g$}_{\alpha\beta}\right)\right> \nonumber \\
&& - g^{\sigma\tau}\left<(\partial_{\mu}\partial_{\sigma}\bg)(\partial_{\nu}\partial_{\tau}\bg)-\frac{1}{2}g_{\mu\nu}g^{\alpha\beta}(\partial_{\alpha}\partial_{\sigma}\bg)(\partial_{\beta}\partial_{\tau}\bg)\right> \nonumber \\
&& + \frac{1}{2}g^{\sigma\tau}\left<\mbox{\boldmath $g$}_{\sigma\mu}(\partial_{\nu}\partial_{\tau}\bg)+\mbox{\boldmath $g$}_{\sigma\nu}(\partial_{\mu}\partial_{\tau}\bg)-g_{\mu\nu}g^{\alpha\beta}\mbox{\boldmath $g$}_{\sigma\alpha}(\partial_{\beta}\partial_{\tau}\bg)\right> \nonumber \\
&& + g^{\sigma\tau}g^{\rho\zeta}\left<\partial_{\zeta}\bg\left(\partial_{\mu}\partial_{\sigma}\bg-\frac{1}{2}\mbox{\boldmath $g$}_{\mu\sigma}\right)\right>\left<\partial_{\rho}\bg\left(\partial_{\nu}\partial_{\tau}\bg-\frac{1}{2}\mbox{\boldmath $g$}_{\nu\tau}\right)\right> \nonumber \\
&& - \frac{1}{2}g_{\mu\nu}g^{\alpha\beta}g^{\sigma\tau}g^{\rho\zeta}\left<\partial_{\zeta}\bg\left(\partial_{\alpha}\partial_{\sigma}\bg-\frac{1}{2}\mbox{\boldmath $g$}_{\alpha\sigma}\right)\right>\left<\partial_{\rho}\bg\left(\partial_{\beta}\partial_{\tau}\bg-\frac{1}{2}\mbox{\boldmath $g$}_{\beta\tau}\right)\right>. \label{etensor}
\end{eqnarray}
In subsequent paragraphs, we consider the physical meaning of Eq.~(\ref{etensor}). In particular, we would like to discuss whether this equation has a 'right' to become the Einstein equation, and then the problem is to know what is the classical field in this abstract information space.

\section{Exponential Family, Hessian Geometry, and Fictitious Energy-Momentum Tensor}

\subsection{Gaussian Distribution}

Let us consider some representative cases in which the present motivation becomes clearer. We first focus on the Gaussian distribution. It has been well-known that the Gaussian distribution naturally leads to the hyperbolic metric~\cite{Amari}, and thus this case strictly holds the vacuum Einstein equation with a negative cosmological constant. Actually, we start with the single-component Gaussian distribution with avarage $\bar{x}$ and standard deviation $\sigma$
\begin{eqnarray}
p(x) = \frac{1}{\sqrt{2\pi}\sigma}\exp\left\{-\frac{(x-\bar{x})^{2}}{2\sigma^{2}}\right\},
\end{eqnarray}
for $X=(-\infty,\infty)$, $-\infty<\bar{x}<\infty$, and $\sigma>0$, and when we take a coordinate \begin{eqnarray}
\bt=(\theta^{1},\theta^{2})=(\bar{x},\sigma),
\end{eqnarray}
we then obtain
\begin{eqnarray}
g_{\mu\nu}d\theta^{\mu}d\theta^{\nu} = \frac{d\bar{x}^{2}+2d\sigma^{2}}{\sigma^{2}}.
\end{eqnarray}
In the Gaussian case, the right hand side of Eq.~(\ref{etensor}) should be reduced to
\begin{eqnarray}
G_{\mu\nu}=-g_{\mu\nu}\Lambda,
\end{eqnarray}
with a negative cosmological constant $\Lambda<0$. In the present formalism, this equation has deeper physical meaning. Substituting Eq.~(\ref{metric2}) into this equation, we obtain
\begin{eqnarray}
G_{\mu\nu} = -\Lambda\left<\partial_{\mu}\bg\partial_{\nu}\bg\right>. \label{fictitious}
\end{eqnarray}
This reminds us that the spectrum $\bg$ or the entropy $S=\left<\bg\right>$ behaves as a scaler field in our classical information space, since the right hand side has structural similarity to the energy-momentum tensor in the Euclidean spacetime. In other words, $\bg$ originally comes from a quantum field theory, but at the same time behaves as a classical field after an appropriate avaraging procedure due to the presence of the bracket. In the Gaussian case, the information space is uniformly bending by the hyperbolic metric, and then it is likely to consider that the effect of data distribution on the information space is described by the simple cosmological constant. In more complex cases where the data distribution is quite nonuniform, it is better to consider that the right hand side of Eq.~(\ref{fictitious}) comes from the energy-momentum tensor of the fictitious matter field into which the data are densely embedded. In order to think this statement is resonable, the field strength should be the amount of physical information in the original microscopic model. Therefore, the entropy, the amount of information, corresponds to the field strength.

The multiple component cases $(x_{1},...,x_{n})$ are also treated in the same manner, and the distribution function and the Fisher metric are given by
\begin{eqnarray}
&& p(x_{1},...,x_{n}) = \frac{1}{\left(\sqrt{2\pi}\sigma\right)^{n}}\exp\left\{-\sum_{i=1}^{n}\frac{(x_{i}-\bar{x}_{i})^{2}}{2\sigma^{2}}\right\}, \label{multiple1} \\
&& g_{\mu\nu}d\theta^{\mu}d\theta^{\nu} = \frac{\sum_{i=1}^{n}d\bar{x}_{i}^{2}+2nd\sigma^{2}}{\sigma^{2}}. \label{multiple2}
\end{eqnarray}
Note that our system has always an Euclidean time axis, when we identify one of avarages $\{\bar{x}_{i}\}$ as the time coordinate in the gravity side. The imaginary time approach is necessary at finite temperatures in the standard quantum field theory. This fact might indicate some coarse-graining by effective finite-temperature in our information space. Actually, microscopic degrees of freedom have been lost in the information space by the avaraging procedure with the probability distribution. Since Eq.~(\ref{multiple1}) should be real, we can not simply apply Wick rotation to one of $\{\bar{x}_{i}\}$. At least for this fact, taking imaginary time axis is quite essential. In this viewpoint, let us remind the Schwarzchild black hole in the Euclidean spacetime coordinate. In this case also, we must take care about periodicity of the imaginary time in order to remove conical singularity, although the removal is only mathematical requirement so that the metric satisfies the Einstein equation, not physical one.

\subsection{Exponential Family}

For further evaluation of Eq.~(\ref{etensor}), we consider the exponential family defined by the following probability distribution
\begin{eqnarray}
p(x;\bt) = \exp\left\{\bt\cdot\mbox{\boldmath $F$}(x)-\psi(\bt)\right\} = \exp\left\{\theta^{\mu}F_{\mu}(x)-\psi(\bt)\right\},
\end{eqnarray}
where $\bt$ is called natural (canonical) parameter. We use continuum representation, but discrete one is also taken in the same manner. The exponential family covers very wide classes of probability distributions. Here we would like to regard the coordinate in the information space as the natural parameter, and thus we use the same notation $\bt$. The function $-\psi(\bt)/\beta$ is usually called as free energy in statistical mechanics. Actually, the Boltzmann distribution is represented by
\begin{eqnarray}
p(x;\bt) = \exp\left\{-\beta E(x;\bt)-\ln Z(\bt)\right\},
\end{eqnarray}
with energy $E$ and partition function $Z$. When the term $-\beta E(x;\bt)$ is decomposed into $\theta^{\mu}$ and $F_{\mu}(x)$, our target model is characterized by the exponential family. At least for classical cases, such decomposition always exists, as we have already seen in Eq.~(\ref{Ising}). Let us present one more example. The single-component Gaussian distribution can be mapped onto the exponential distribution as follows
\begin{eqnarray}
p(x) = \frac{1}{\sqrt{2\pi}\sigma}\exp\left\{-\frac{(x-\bar{x})^{2}}{2\sigma^{2}}\right\} = \exp\left\{-\ln\left(\sqrt{2\pi}\sigma\right)-\frac{x^{2}}{2\sigma^{2}}+\frac{x\bar{x}}{\sigma^{2}}-\frac{\bar{x}^{2}}{2\sigma^{2}}\right\},
\end{eqnarray}
where we find that $\mbox{\boldmath $F$}$, $\bt$, and $\psi(\bt)$ are defined by
\begin{eqnarray}
\mbox{\boldmath $F$}&=&\left(F_{1},F_{2}\right)=\left(x,x^{2}\right), \\
\bt&=&\left(\theta^{1},\theta^{2}\right)=\left(\frac{\bar{x}}{\sigma^{2}},-\frac{1}{2\sigma^{2}}\right),
\end{eqnarray}
and
\begin{eqnarray}
\psi(\bt)=\ln\left(\sqrt{2\pi}\sigma\right)+\frac{\bar{x}^{2}}{2\sigma^{2}}=\frac{1}{2}\ln\left(-\frac{\pi}{\theta^{2}}\right)-\frac{(\theta^{1})^{2}}{4\theta^{2}}.
\end{eqnarray}

In the exponential family, the spectrum $\bg$ is given by
\begin{eqnarray}
\bg=-\ln p(x;\bt)=\psi(\bt)-\theta^{\nu}F_{\nu}(x), \label{gamma2}
\end{eqnarray}
and thus $\psi(\bt)$ is directly related to the information $\bg$ of the original model. The first and second derivatives of $\bg$ are given by
\begin{eqnarray}
\partial_{\mu}\bg &=& \partial_{\mu}\psi(\bt)-F_{\mu}(x), \label{partial1} \\
\partial_{\mu}\partial_{\nu}\bg &=& \partial_{\mu}\partial_{\nu}\psi(\bt). \label{partial2}
\end{eqnarray}
We see that the $x$ dependence of $\bg$ vanishes by the second derivative. After the second derivative, $\psi(\bt)$ can be identified with $\bg$. Then, the Fisher metric is represented by
\begin{eqnarray}
g_{\mu\nu} &=& \left<\partial_{\mu}\partial_{\nu}\bg\right>=\partial_{\mu}\partial_{\nu}\psi(\bt), \label{rep1} \\
g_{\mu\nu} &=& \left<\partial_{\mu}\bg\partial_{\nu}\bg\right>=\partial_{\mu}\psi(\bt)\partial_{\nu}\psi(\bt)-\left(\left<F_{\mu}\right>\partial_{\nu}+\left<F_{\nu}\right>\partial_{\mu}\right)\psi(\bt)+\left<F_{\mu}F_{\nu}\right>. \label{rep2}
\end{eqnarray}
The first equation is so-called Hessian structure. It would be common for string theorists that $\psi(\bt)$ corresponds to a real version of the K\"{a}hler potential in the complex manifold theory. In addition, the expectation value of Eq.~(\ref{partial1}) is given by
\begin{eqnarray}
\left<F_{\mu}\right>=\partial_{\mu}\psi(\bt). \label{f1}
\end{eqnarray}
Substituting this with Eq.~(\ref{rep2}), we obtain
\begin{eqnarray}
g_{\mu\nu}=\left<F_{\mu}F_{\nu}\right>-\partial_{\mu}\psi(\bt)\partial_{\nu}\psi(\bt)=\left<F_{\mu}F_{\nu}\right>-\left<F_{\mu}\right>\left<F_{\nu}\right>.
\end{eqnarray}
Thus, the metric is a covariance matrix of $F_{\mu}$. We also differentiate $g_{\mu\nu}=\left<\partial_{\mu}\bg\partial_{\nu}\bg\right>$ by $\theta^{\lambda}$, and obtain
\begin{eqnarray}
\partial_{\lambda}g_{\mu\nu}=-\left<\partial_{\lambda}\bg\partial_{\mu}\bg\partial_{\nu}\bg\right>+\left<\left(\partial_{\lambda}\partial_{\mu}\bg\right)\partial_{\nu}\bg\right>+\left<\partial_{\mu}\bg\left(\partial_{\lambda}\partial_{\nu}\bg\right)\right>=\partial_{\lambda}\partial_{\mu}\partial_{\nu}\psi(\bt).
\end{eqnarray}
For the exponential family, this equation means
\begin{eqnarray}
T_{\lambda\mu\nu}=\left<\partial_{\lambda}\bg\partial_{\mu}\bg\partial_{\nu}\bg\right>=-\partial_{\lambda}\partial_{\mu}\partial_{\nu}\psi(\bt).
\end{eqnarray}
This tensor plays a central role in the Hessian geometry, since the tensor is colosely related to the Christoffel symbol as
\begin{eqnarray}
\Gamma^{\lambda}_{\;\mu\nu} = -\frac{1}{2}g^{\lambda\tau}T_{\tau\mu\nu}.
\end{eqnarray}

By using the expomential family, we transform Eq.~(\ref{etensor}) into a more compact form. The first term of Eq.~(\ref{etensor}) is then given by
\begin{eqnarray}
\left<\left(\mbox{\boldmath $\omega$}+\bp\right)\left(\partial_{\mu}\partial_{\nu}\bg-\frac{1}{2}g_{\mu\nu}g^{\alpha\beta}\partial_{\alpha}\partial_{\beta}\bg\right)\right> = \frac{1}{2}D\left(1-\frac{1}{2}D\right)g_{\mu\nu}.
\end{eqnarray}
The second term is given by
\begin{eqnarray}
&& \frac{1}{2}\left<\left(\bp+g^{\sigma\tau}\partial_{\sigma}\partial_{\tau}\bg\right)\left(\mbox{\boldmath $g$}_{\mu\nu}-\frac{1}{2}g_{\mu\nu}g^{\alpha\beta}\mbox{\boldmath $g$}_{\alpha\beta}\right)\right> \nonumber \\
&& \;\;\;\; = \frac{1}{4}g^{\sigma\tau}g^{\rho\zeta}\left<\left(\partial_{\rho}\bg\right)\mbox{\boldmath $g$}_{\sigma\tau}\right>\left<\partial_{\zeta}\bg\left(\mbox{\boldmath $g$}_{\mu\nu}-\frac{1}{2}g_{\mu\nu}g^{\alpha\beta}\mbox{\boldmath $g$}_{\alpha\beta}\right)\right> + \frac{1}{2}D\left(1-\frac{1}{2}D\right)g_{\mu\nu}.
\end{eqnarray}
The third term is given by
\begin{eqnarray}
g^{\sigma\tau}\left<(\partial_{\mu}\partial_{\sigma}\bg)(\partial_{\nu}\partial_{\tau}\bg)-\frac{1}{2}g_{\mu\nu}g^{\alpha\beta}(\partial_{\alpha}\partial_{\sigma}\bg)(\partial_{\beta}\partial_{\tau}\bg)\right> = \left(1-\frac{1}{2}D\right)g_{\mu\nu}.
\end{eqnarray}
The fourth term is given by
\begin{eqnarray}
\frac{1}{2}g^{\sigma\tau}\left<\mbox{\boldmath $g$}_{\sigma\mu}(\partial_{\nu}\partial_{\tau}\bg)+\mbox{\boldmath $g$}_{\sigma\nu}(\partial_{\mu}\partial_{\tau}\bg)-g_{\mu\nu}g^{\alpha\beta}\mbox{\boldmath $g$}_{\sigma\alpha}(\partial_{\beta}\partial_{\tau}\bg)\right> = \left(1-\frac{1}{2}D\right)g_{\mu\nu}.
\end{eqnarray}
Up to now, we have found that the terms proportional to $(1-D/2)g_{\mu\nu}$ cancel out. Summarizing, we obtain the Einstein tensor for the exponential family as
\begin{eqnarray}
G_{\mu\nu} &=& -\frac{1}{4}g^{\sigma\tau}g^{\rho\zeta}\left<\left(\partial_{\rho}\bg\right)\mbox{\boldmath $g$}_{\sigma\tau}\right>\left<\partial_{\zeta}\bg\left(\mbox{\boldmath $g$}_{\mu\nu}-\frac{1}{2}g_{\mu\nu}g^{\alpha\beta}\mbox{\boldmath $g$}_{\alpha\beta}\right)\right> \nonumber \\
&& + \frac{1}{4}g^{\sigma\tau}g^{\rho\zeta}\left<\left(\partial_{\zeta}\bg\right)\mbox{\boldmath $g$}_{\mu\sigma}\right>\left<\left(\partial_{\rho}\bg\right)\mbox{\boldmath $g$}_{\nu\tau}\right> \nonumber \\
&& - \frac{1}{8}g_{\mu\nu}g^{\alpha\beta}g^{\sigma\tau}g^{\rho\zeta}\left<\left(\partial_{\zeta}\bg\right)\mbox{\boldmath $g$}_{\alpha\sigma}\right>\left<\left(\partial_{\rho}\bg\right)\mbox{\boldmath $g$}_{\beta\tau}\right> \nonumber \\
&=& X_{\mu\nu} - \frac{1}{2}g_{\mu\nu}g^{\alpha\beta}X_{\alpha\beta},
\end{eqnarray}
where
\begin{eqnarray}
X_{\mu\nu} &=& \frac{1}{4}g^{\sigma\tau}g^{\rho\zeta}\left( T_{\zeta\mu\sigma}T_{\rho\nu\tau}-T_{\rho\sigma\tau}T_{\zeta\mu\nu} \right) \label{xr1} \\
&=& \frac{1}{4}g^{\sigma\tau}g^{\rho\zeta}\left( \left<\partial_{\zeta}\bg\partial_{\mu}\bg\partial_{\sigma}\bg\right>\left<\partial_{\rho}\bg\partial_{\nu}\bg\partial_{\tau}\bg\right> - \left<\partial_{\rho}\bg\partial_{\sigma}\bg\partial_{\tau}\bg\right>\left<\partial_{\zeta}\bg\partial_{\mu}\bg\partial_{\nu}\bg\right> \right). \label{xr2}
\end{eqnarray}

\subsection{Derivation of Energy-Momentum Tensor}

We would like to transform $G_{\mu\nu}$ for the exponential family into a new form in which we can see that the potential $\psi(\bt)$ actually behaves as a classical field in the information space. For this purpose, we pick up terms associated with $\partial_{\mu}\psi$ and $\partial_{\nu}\psi$ in Eq.~(\ref{xr2}), and then $X_{\mu\nu}$ is evaluated as
\begin{eqnarray}
X_{\mu\nu} &=& \frac{1}{4}D\partial_{\mu}\psi\partial_{\nu}\psi - \frac{1}{4}g^{\rho\tau}\partial_{\mu}\psi\left<F_{\nu}\mbox{\boldmath $g$}_{\rho\tau}\right> - \frac{1}{4}g^{\sigma\zeta}\partial_{\nu}\psi\left<F_{\mu}\mbox{\boldmath $g$}_{\zeta\sigma}\right> \nonumber \\
&& + \frac{1}{4}g^{\sigma\tau}g^{\rho\zeta}\left<F_{\mu}\mbox{\boldmath $g$}_{\zeta\sigma}\right>\left<F_{\nu}\mbox{\boldmath $g$}_{\rho\tau}\right> \nonumber \\
&& + \frac{1}{4}g^{\sigma\tau}g^{\rho\zeta}\left<\partial_{\rho}\bg\partial_{\sigma}\bg\partial_{\tau}\bg\right>\left(\left<F_{\nu}\partial_{\zeta}\bg\right>\partial_{\mu}\psi+\left<F_{\mu}\partial_{\zeta}\bg\right>\partial_{\nu}\psi\right) \nonumber \\
&& - \frac{1}{4}g^{\sigma\tau}g^{\rho\zeta}\left<\partial_{\rho}\bg\partial_{\sigma}\bg\partial_{\tau}\bg\right>\left<F_{\mu}F_{\nu}\partial_{\zeta}\bg\right>.
\end{eqnarray}
Here, using the relation
\begin{eqnarray}
\left<F_{\nu}\partial_{\zeta}\bg\right>=\partial_{\zeta}\psi(\bt)\left<F_{\nu}\right>-\left<F_{\zeta}F_{\nu}\right>=-g_{\nu\zeta},
\end{eqnarray}
we find
\begin{eqnarray}
X_{\mu\nu} &=& \frac{1}{4}g^{\sigma\tau}g^{\rho\zeta}\left(\left<F_{\mu}\mbox{\boldmath $g$}_{\zeta\sigma}\right>\left<F_{\nu}\mbox{\boldmath $g$}_{\rho\tau}\right> - \left<F_{\mu}\right>g_{\zeta\sigma}\left<F_{\nu}\right>g_{\rho\tau}\right) \nonumber \\
&& - \frac{1}{4}g^{\sigma\tau}g^{\rho\zeta}\left<\partial_{\rho}\bg\partial_{\sigma}\bg\partial_{\tau}\bg\right>\left<F_{\mu}F_{\nu}\partial_{\zeta}\bg\right>. \label{fl}
\end{eqnarray}
The Equations~(\ref{xr1}), (\ref{xr2}) and (\ref{fl}) are three typical representations that characterize the geometry of this system.

We particulary focus on the first term of Eq.~(\ref{fl}) in order to take some approximation, since the second term is approximately given by $\left<F_{\mu}F_{\nu}\partial_{\zeta}\bg\right>\sim\left<F_{\mu}F_{\nu}\right>\left<\partial_{\zeta}\bg\right>=0$. We expand $\mbox{\boldmath $g$}_{\mu\nu}(x;\bt)$ near the avarage of $x$, $\left<x\right>=\bar{x}$. Then, we have
\begin{eqnarray}
\left<F_{\mu}\mbox{\boldmath $g$}_{\zeta\sigma}\right> &=& \left<F_{\mu}\left(\mbox{\boldmath $g$}_{\zeta\sigma}(\bar{x};\bt)+\mbox{\boldmath $\dot{g}$}_{\zeta\sigma}(\bar{x};\bt)(x-\bar{x})+\frac{1}{2}\mbox{\boldmath $\ddot{g}$}_{\zeta\sigma}(\bar{x};\bt)(x-\bar{x})^{2}+\cdots\right)\right> \nonumber \\
&=& \mbox{\boldmath $g$}_{\zeta\sigma}(\bar{x};\bt)\partial_{\mu}\psi + \mbox{\boldmath $\dot{g}$}_{\zeta\sigma}(\bar{x};\bt)\left<F_{\mu}(x-\bar{x})\right> + \frac{1}{2}\mbox{\boldmath $\ddot{g}$}_{\zeta\sigma}(\bar{x};\bt)\left<F_{\mu}(x-\bar{x})^{2}\right> + \cdots,
\end{eqnarray}
where the dot represents differentiation of the metric by $x$
\begin{eqnarray}
\mbox{\boldmath $\dot{g}$}_{\zeta\sigma}(\bar{x};\bt)=\lim_{x\rightarrow\bar{x}}\frac{\partial}{\partial x}\mbox{\boldmath $g$}_{\zeta\sigma}(x;\bt).
\end{eqnarray}
Then, we obtain
\begin{eqnarray}
X_{\mu\nu} &=& \frac{1}{4}g^{\sigma\tau}g^{\rho\zeta}\left(\mbox{\boldmath $g$}_{\zeta\sigma}(\bar{x};\bt)\mbox{\boldmath $g$}_{\rho\tau}(\bar{x};\bt)-g_{\zeta\sigma}g_{\rho\tau}\right)\partial_{\mu}\psi\partial_{\nu}\psi \nonumber \\
&& + \frac{1}{4}g^{\sigma\tau}g^{\rho\zeta}\left(\mbox{\boldmath $\dot{g}$}_{\zeta\sigma}(\bar{x};\bt)\left<F_{\mu}(x-\bar{x})\right>+\frac{1}{2}\mbox{\boldmath $\ddot{g}$}_{\zeta\sigma}(\bar{x};\bt)\left<F_{\mu}(x-\bar{x})^{2}\right>+\cdots\right)\mbox{\boldmath $g$}_{\rho\tau}(\bar{x};\bt)\partial_{\nu}\psi \nonumber \\
&& + \frac{1}{4}g^{\sigma\tau}g^{\rho\zeta}\mbox{\boldmath $g$}_{\zeta\sigma}(\bar{x};\bt)\partial_{\mu}\psi\left(\mbox{\boldmath $\dot{g}$}_{\rho\tau}(\bar{x};\bt)\left<F_{\mu}(x-\bar{x})\right>+\frac{1}{2}\mbox{\boldmath $\ddot{g}$}_{\rho\tau}(\bar{x};\bt)\left<F_{\mu}(x-\bar{x})^{2}\right>+\cdots\right) \nonumber \\
&& + \cdots.
\end{eqnarray}
The expansion terminates up to the fourth-order terms in the normal distribution case. If we identify $\mbox{\boldmath $g$}_{\mu\nu}(\left<x\right>;\bt)=\mbox{\boldmath $g$}_{\mu\nu}(\bar{x};\bt)$ with $g_{\mu\nu}=\left<\mbox{\boldmath $g$}_{\mu\nu}(x;\bt)\right>$ and we introduce the following approximation
\begin{eqnarray}
\left<F_{\mu}\left(x-\bar{x}\right)^{n}\right> \simeq \partial_{\mu}\psi\left<(x-\bar{x})^{n}\right>,
\end{eqnarray}
then we obtain
\begin{eqnarray}
X_{\mu\nu} = \kappa\partial_{\mu}\psi(\bt)\partial_{\nu}\psi(\bt),
\end{eqnarray}
with a constant $\kappa$. For the normal distribution, we have 
\begin{eqnarray}
\left<(x-\bar{x})^{n}\right>=\sigma^{n}\prod_{k=1}^{n/2}(2k-1),
\end{eqnarray}
and $\kappa$ can be represented as
\begin{eqnarray}
\kappa \simeq \frac{1}{4}\sigma^{2}g^{\sigma\zeta}\mbox{\boldmath $\ddot{g}$}_{\zeta\sigma}(\bar{x};\bt)+\frac{1}{16}\sigma^{4}g^{\sigma\tau}g^{\rho\zeta}\mbox{\boldmath $\ddot{g}$}_{\zeta\sigma}(\bar{x};\bt)\mbox{\boldmath $\ddot{g}$}_{\rho\tau}(\bar{x};\bt)+\cdots.
\end{eqnarray}
Therefore, the K\"{a}hler potential $\psi(\bt)$ does not only corresponds to the entropy value in the microscopic-model side, but really behaves as a classical scalar field in the information space side. The parameter $\kappa$ is dimensionless, since we do not introduce any energy scale and all of the parameters $\bt$ are normalized by the temperature.

Let us define the effective Lagrangian for the scaler field $\psi$ as
\begin{eqnarray}
L = \frac{1}{2}g^{\alpha\beta}\partial_{\alpha}\psi(\bt)\partial_{\beta}\psi(\bt),
\end{eqnarray}
and then the energy-momentum tensor can be represented as
\begin{eqnarray}
T_{\mu\nu} = g_{\mu\sigma}\frac{\partial L}{\partial(\partial_{\sigma}\psi)}\partial_{\nu}\psi - g_{\mu\nu}L = \frac{1}{\kappa}\left(X_{\mu\nu}-\frac{1}{2}g_{\mu\nu}g^{\alpha\beta}X_{\alpha\beta}\right).
\end{eqnarray}
Finally, we reach the Einstein equation
\begin{eqnarray}
G_{\mu\nu}=\kappa T_{\mu\nu}.
\end{eqnarray}

\section{Discussion and Summary}

Based on the present result, we discuss related topics. The first one is about quantization of gravity. My massage is that the quantum gravity theory may not come from quantization of the Einstein gravity, rather the Einstein equation can be regarded as the equation of the coarse-grained quantum state. Of course, the present approach is not unique one, but still gives us quite important information for the construction of the quantum gravity theory. The present result supports the previous and recent development in which the gravity is a kind of entropic force~\cite{Padmanabhan,Jacobson,Eling,Verlinde}.

In the standard information geometry, the Kullback-Leiber divergence is used for statistical inference. We infer the appropriate probability distribution from a given data set characterized by $\bt$. Then, the precision of the inference is bounded by the inverse of the Fisher metric. The classical space side only contains the natural canonical parameter $\bt$ and not $x$, and thus this is not beyond the inference. In this case also, the most elementary data are probability distributions, and the observed data are regarded as their coarse-grained data set.

A future problem to be resolved is to understand causal structure of our information spacetime. To do this, it is necessary to examine time evolution of the microscopic system. We expect that the causal structure is preserved in the classical side, even though we start from quantum field where the non-locality may violate the causal nature. The point is that we mainly consider the entropy, not the quantum state itself. Now, we once terminate time evolution of the quantum state in order to determine the property of the entropy at a particular time. Thus, the time coordinate is special in comparison with the other coordinates. In this case, the classical side looks like a 'laminated' body, and each layer is connected smoothly. Thus, the dynamic exchange of these layers does not occur.

As already mentioned in the previous paragraphs, the present approach naturally represents quantum-classical correspondence, when we start from quantum field theory in the microscopic side. This is a new type of correspondence, when we compare it with other famous correspondences, such as the Suzuki-Trotter decomposition in statistical and condensed matter physics~\cite{Suzuki} and the anti-de Sitter space / conformal field theory (AdS/CFT) correspondence in string theory~\cite{Maldacena}. Their most remarkable difference is holographic nature of each correspondence. Usually, holographic theories contain the extra dimension controlling an energy scale of the original quantum model. In this viewpoint, the theory is characterized by the renormalization concept. In the information-geometrical approach, however, the extra dimension seems to be formally missing. This is because the parameter $\bt$ automatically contains temperature, and then the change in $\bt$ means indirect control of accesible lowest energy scale. Therefore, it is a very important future work to clarify similarity and difference among them. A hint to resolve this problem is to examine more about the Gaussian distribution. In the Gaussian case, the variance $\sigma^{2}$ plays a similar role on temperature in the Boltzmann distribution. Then, the variance characterizes the energy (or length) scale. Furthermore the von Neumann entropy is given by
\begin{eqnarray}
S=\left<\bg\right>=\ln\left(\sqrt{2\pi}\sigma\right).
\end{eqnarray}
Thus, the total amount of information $S$ is a logarithmic function of the variance. This feature seems to be comparable to the scaling properties of the entanglement entropy in CFT~\cite{Holzhey,Calabrese,Tagliacozzo,Pollmann,Huang} (Fixing the $\sigma$ value indicates that the accesible data sets are finite, leading to the calculation of the partial density matrix and the entanglement entropy). As already mentioned, the Fisher metric for the Gaussian distribution becomes AdS. In a view point of AdS/CFT, the problem is then how the Gaussian distribution comes from one-dimensional quantum critical theories, and this is on-going work.

Summarizing, we have derived the classical Einstein equation from the Fisher information metric defined by the microscopic statistical data. In this formalism, the Einstein equation is a kind of the equation of the coarse-grained quantum states. The present approach is based on information-geometrical techniques, and thus may not be familiar for many researchers. However, the author believes that more flexible use of this kind of approaches opens new doors of physics of quantum-classical correspondence and micro-macro duality.


\begin{thebibliography}{99}

\bibitem{Solodukhin}
Sergey N. Solodukhin, Living Rev. Relativity, {\bf 14}, 8 (2011).
\bibitem{Matsueda1}
H. Matsueda, arXiv:1112.5566.
\bibitem{Amari}
Shun-ichi Amari and Hiroshi Nagaoka, "Methods of Information Geometry", Oxford (2000).
\bibitem{Cencov}
N. N. \v{C}encov, "Statistical Decision Rules and Optimal Inference", vol. 53, translations of mathematical monographs, AMS (1982) [original: Russian (1972)].
\bibitem{Shima}
Hirohiko Shima, "Hessian Goemetry", Shokabo, Tokyo (2001).
\bibitem{Matsueda2}
H. Matsueda, arXiv:1208.5103.
\bibitem{Padmanabhan}
T. Padmanabhan, Class. Quant. Grav. {\bf 19}, 5387 (2002).
\bibitem{Jacobson}
T. Jacobson, Phys. Rev. Lett. {\bf 75}, 1260 (1995).
\bibitem{Eling}
C. Eling, R. Guedens, and T. Jacobson, Phys. Rev. Lett. {\bf 96}, 121301 (2006).
\bibitem{Verlinde}
E. P. Verlinde, JHEP {\bf 1104}, 029 (2011).

\bibitem{Suzuki}
M. Suzuki, Prog. Theor. Phys. {\bf 56}, 1454 (1976).
\bibitem{Maldacena}
J. M. Maldacena, Adv. Theor. Math. Phys. {\bf 2}, 231 (1998).

\bibitem{Holzhey}
Christoph Holzhey, Finn Larsen, Frank Wilczek, Nucl. Phys. B {\bf 424}, 443 (1994).
\bibitem{Calabrese}
P. Calabrese and J. Cardy, J. Stat. Mech. 0406 (2004) P002 [note added: arXiv:hep-th/0405152].
\bibitem{Tagliacozzo}
L. Tagliacozzo, Thiago. R. de Oliveira, S. Iblisdir, amd J. I. Latorre, Phys. Rev. B {\bf 78}, 024410 (2008).
\bibitem{Pollmann}
Frank Pollmann, Subroto Mukerjee, Ari M. Turner, and Joel E. Moore, Phys. Rev. Lett. {\bf 102}, 255701 (2009).
\bibitem{Huang}
Ching-Yu Huang and Feng-Li Lin, Phys. Rev. A {\bf 81}, 032304 (2010).


\end{thebibliography}
\end{document}